\documentclass[preprint2]{proto}
\usepackage{times}
\usepackage{xcolor}
\usepackage{graphicx}
\usepackage{wasysym}
\usepackage{float}
\usepackage{dblfloatfix}
\usepackage{tabularx}
\usepackage{rotating}
\usepackage{lscape}
\usepackage{soul}
\usepackage{url}

\graphicspath{{./}{Figures/}}

\begin{document}

\newcommand\rev[1]{\textcolor{black}{ #1}}

\title{\textbf{\LARGE COMET SCIENCE WITH GROUND BASED AND SPACE BASED SURVEYS IN THE NEW MILLENNIUM}}

\author {\textbf{\large J. M. Bauer}}
\affil{\small\em Dept. of Astronomy, Univ. of Maryland, College Park, MD, USA, 20742-2421}

\author{\textbf{Y. R. Fern\'{a}ndez}}
\affil{\small\em Dept. of Physics and Florida Space Inst., Univ. of Central Florida, 4000 Central Florida Blvd., Orlando, FL, USA 32816-2385}

\author {\textbf{\large S. Protopapa}}
\affil{\small\em Southwest Research Institute, 1050 Walnut Street, Suite 300, Boulder, CO, USA 80302}

\author {\textbf{\large L. M. Woodney}}
\affil{\small\em California State University, San Bernardino 5500 University Parkway San Bernardino, CA, USA 92407}

\begin{abstract}

\begin{list}{ } {\rightmargin 0.4in}
\baselineskip = 11pt
\parindent=1pc
{\small 
We summarize the \rev{comet} science provided by surveys. This includes surveys where the detections of comets are an advantageous benefit but were not part of the survey’s original intent, as well as some pointed surveys where comet science was the goal. Many of the surveys are made using astrophysical and heliophysics assets. The surveys in our scope include those using ground-based as well as space-based telescope facilities. Emphasis is placed on current or recent surveys, and science that has resulted since the publication of Comets II, though key advancements made by earlier surveys (e.g. IRAS, COBE, NEAT, etc.) will be mentioned. The proportionally greater number of discoveries of comets by surveys have yielded in turn larger samples of \rev{comet} populations and sub-populations for study, resulting in better defined evolutionary trends. While providing an array of remarkable discoveries, most of the survey data has been only cursorily investigated. It is clear that continuing to fund ground- and space-based surveys of large numbers of comets is vital if we are to address science goals that can give us a population-wide picture of \rev{comet} properties.
\\~\\~\\~}
\end{list}
\end{abstract}  

\section{\textbf{INTRODUCTION}}
\label{sec:intro}




Over the span of the last decade and a half, a more automated approach to the analysis of data has become common. This is in part owing to the arrival of data sets which are so large \rev{that each of the observations is not practically analyzed by human interaction, but rather is conducted by automated pipeline.} Analysis routines are prototyped, tested, and applied to larger datasets, while outliers and diagnostic triggers indicate where special circumstances apply, and further manipulation, or rejection, of the data are required. Much of this has been driven  by the advent of the vast quantities of data provided by automated sky surveys. Additionally specialized data sets that are the product of targeted observations now have a certain expectation of providing statistically significant samples large enough for outliers to be identified and trends to be discerned.

Previous generation surveys generally had relatively small sample sizes. Many of these surveys, demonstrably the space-based surveys, made robust discoveries with these smaller samples. \cite{1998ApJ...496..971L} found temperature excess in dust comae from observations of five comets at perihelion distances $\sim 1$au obtained by the Cosmic Microwave Background Explorer (COBE).  R\"{o}ntgen Satellite (ROSAT) observations of six \citep{1997Sci...277.1625D}, and later eleven (c.f. \citealt{2004come.book..631L}), comets revealed charge exchange between highly charged heavy ions in the solar wind and cometary neutrals dominated cometary X-ray emissions. A subsequent survey by \cite{2007A&A...469.1183B} of eight comets with the Chandra observatory found that the characteristics of observed X-ray spectra mainly reflect the state of the local solar wind. The Infrared Astronomy Satellite (IRAS) mission data provided the first thermal dust trail measurements from eight identified comets \citep{1992Icar...95..180S}. Such surveys had large impacts on the cometary field, but did not employ the more automated large-sample approaches, such as with astroinformatic techniques (c.f. \citealt{2009astro2010P...6B}), now utilized for larger survey samples.

For the definition of survey within these pages, we include sample sizes of 20 or greater, owing to the conditions that even for simple statistical correlation tests, sample sizes $\sim 20$ or greater are required to achieve $95\%$ confidence values even for strong correlations (c.f. \citealt{2000BonnetWright}). Here we concentrate on classical comet populations (long-period and short-period comets; \rev{LPCs and SPCs, respectively}) and their dynamically defined sub-classifications (c.f. \citealt{1996ASPC..107..173L}). \rev{SPCs are defined to have orbital periods $< 200$ years, and LPCs having orbit periods $\gtrsim 200$ years. Jupiter Family Comets (JFCs), for example, are a sub-class of SPCs with orbital periods $\lesssim 20$ years and pro-grade low orbital inclinations $\lesssim  40^{\circ}$, while dynamically new comets are a sub-class of LPCs with original orbital semi-major axis values $\gtrsim 10^4$ au. Generally
speaking, the source of LPCs is the Oort cloud while the Kuiper belt feeds the population of SPCs. Notably Halley-type comets (HTCs) have historically often been lumped with SPCs, but most of them are likely to be highly-evolved (in the dynamics sense) objects from the Oort Cloud. Thus they are more closely related to the LPCs. There are also different opinions on the meaning of the term 'Oort Cloud comet'; e.g., it may only include dynamically new comets, or it may include all LPCs and HTCs that were in the Oort Cloud any time in the past.}

Measurements of large populations from single platforms and the same, or similar, instrumentation provide a basis for comparative samples, in contrast with compilations (c.f. \rev{ \citealt{AHearn1995}}, \citealt{2020NatAs...4..930L} and \citealt{AHearn2012}). Such samples of cometary physical properties may be targeted, such as narrow-band filter surveys (cf. \citealt{2004come.book..449S}) or spectroscopic surveys (cf. \citealt{2016Icar..278..301D}), or serendipitous observations, such as the data obtained with many ground-based or space-based sky surveys\footnote{Here {\it sky survey} refers to a survey which covers regions of the inertial frame, or background, sky with target coordinates fixed in the equatorial, ecliptic, or galactic coordinate frames, as opposed to moving targets (or solar system objects).}. These two categories are significantly different in the selection of the objects observed, and how representative the samples are of the background populations.

In the first case of targeted samples, known solar-system object targets are selected based on their optical brightness, and were discovered often by \rev{sky} surveys. They are often observed at preferred geometries (opposition, for example, for ground based telescopic surveys) and detected while they are most active. As such, there are potential selection biases in the sample that may skew projection of behavior or physical properties of the base population. For  targeted observations the observing time can be selected to sample the points through a comet's orbit where the expected levels of activity are best matched to the physical property of interest. For example, optical surveys of comets at aphelion may provide more accurate absolute magnitude values of the nucleus, leading to better derived reflectances if the size of the body is known. Alternatively, a more comprehensive inventory of gas species may be derived at perihelion where so-called hyper-volatiles and water-related species are released, and following a comet through its orbit may reveal when particular species dominate the activity.

\subsection{Survey Discoveries of Comets}
\label{sec:discoverystats}

Prior to the 1990s, comets were generally discovered either in large photographic plate exposures or by individuals that visually scanned the sky, often employing specialized telescopes or binoculars with fast optics. In the late 1980s and early 1990s, digital cameras began to be employed in regular searches of the sky for solar system objects (cf. \citealt{1991LPICo.765..191S}) with a handful of early comet discoveries. In addition, astrophysical sky surveys were conceived to identify transient behavior, like supernova events, in extra-solar-system objects. With the advent of the earliest digital sky surveys, the automated surveys began to make significant contributions in the number of comet discoveries in the mid-1990s. These foundational surveys employed Charge-Coupled Device (CCD) cameras with fields of view that by today's standards would be quite modest, on the order of a degree on a side (c.f. \citealt{1999AJ....117.1616P}), and rapidly began to outpace other means of discovery.

The efforts to detect solar system objects have been largely driven by the intent to discover and characterize Near Earth Objects (NEOs). Observing cadences, pointing strategy, and approaches to analysis were therefore optimized or prioritized towards these NEO-related goals. \rev{These efforts have been remarkably effective, and discovery of more than 83\% of the known NEOs has been the result of these efforts} \citep{2019AcAau.156..394L}.  As a means of discovering comets, the NEO search programs have also been effective, not only with the comets that are a component of the NEO population, but also with more distant comets.

As of September 30, 2021, approximately 3586 comets had been discovered, as registered by the Jet Propulsion Laboratory (JPL) small bodies database.\footnote{ https://ssd.jpl.nasa.gov/tools/sbdb$\_$query.html} A summary of leading discovery platforms is provided in Table \ref{tab:discsumm}. According to the database, 41$\%$ of the comet discoveries listed were discovered by the Solar and Heliospheric Observatory (SOHO; with the SWAN and LASCO instruments\footnote{SWAN: the Solar Wind ANisotropy experiment and LASCO: the Large Angle and Spectrometric COronagraph instrument}) or the Solar Terrestrial Relations Observatory (STEREO) spacecraft. In total, including these surveys, over 71\% of comet discoveries up to October 2021 have been made by sky surveys.  It is worth noting that the Minor Planet Center count ($\sim$4430 comets as of September 30, 2021), and future counts in the near term, are likely to be higher, as \rev{a significant remainder of the data from the SOHO spacecraft have yet to be processed and the JPL number includes only those comets} that have been observed by other non-solar-observing platforms in addition.  Figure 1 shows the annual number of discoveries and observations of objects by sky surveys reported to the MPC and listed in the JPL database. The drop-off in the discoveries near 2010 coincides with the curtailment of sun-pointing spacecraft survey data by the MPC, which has been recently resumed \citep{2020MPEC....W...Y19}, though has not yet  encompassed the multi-year backlog \citep{2017RSPTA.37560257B}.

\begin{table}[h]
\caption{Comet Discoveries by Selected Sky Surveys \label{tab:discsumm}}
\begin{tabularx}{\columnwidth}{lcccc}
\noalign{\vskip 5 pt}
\hline
\hline\noalign{\vskip 5 pt}
Survey & Location$^a$ & SPCs$^{b}$ & LPCs$^{b}$ & Total \\
\hline
 Catalina    & G96, 703, I52 &200& 193 & 393 \\
 Pan-STARRS  & F51, F52      &112& 142 & 254 \\
 LINEAR      & 704           &91 & 128 & 219 \\
 NEAT        & 566, 644      &39 & 16  & 55 \\
 ATLAS       & T05, T07, T08 &16  & 35  & 51 \\
 NEOWISE& C51           & 16 & 23  & 39 \\
 Spacewatch  &  291, 691     &16 & 12  & 28 \\
 LONEOS      & 699           &17 & 5   & 22 \\
 ZTF/PTF     & I41           &2  & 16   & 18 \\
\hline
 SOHO/SWAN$^{c}$ & 249 & 12 & 1466 & 1478 \\
 STEREO$^{c}$ & C49 & 1 & 8 & 9\\
\end{tabularx}
\tablenotetext{a}{The Minor Planet Center Observatory Code contributors.}
\tablenotetext{b}{Short-period comets (SPCs) with orbital periods $< 200$ years \\ and Long-period comets (LPCs) with orbital periods $\geq 200$ years.}
\tablenotetext{c}{Sun-looking survey total.}
\tablecomments{Note that the count of SOHO-discovered comets include \\ only those contributions with additional non-SOHO observations (see text). \\}
\end{table}

\subsection{Survey Observations}
\label{sec:surveyobsstats}

\begin{figure*}[t]
\begin{center}
\includegraphics[width=7.5in]{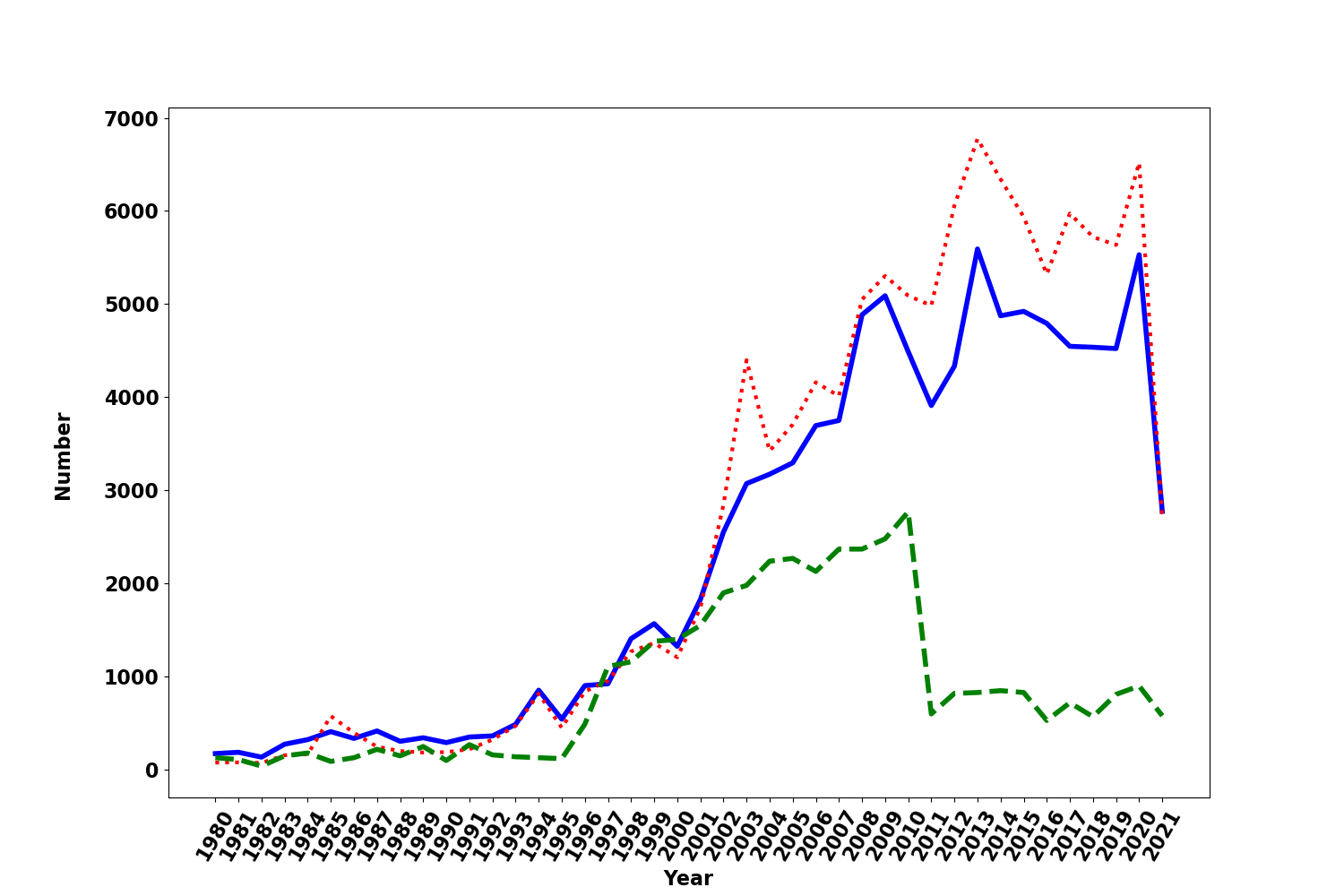}
\caption{Discoveries and reported observations of comets per year, 1980-2021. Green dashed line indicates discoveries per year (multiplied by 10) as listed in the JPL small bodies database (but see text for explanation of the 2010 drop off). Blue solid line indicates the number of comets with observations reported to the MPC. Red dotted line indicates the number of observations reported to the MPC (divided by 10). }
\label{fig:Xmas}
\end{center}
\end{figure*}

Along with a marked increase in the number of comet discoveries brought through ground-based surveys, the number of observations of comets has increased as well. Figure \ref{fig:Xmas} shows that the number of observations closely tracks the number of objects observed by the surveys.\footnote{Note that the drop in 2021 in Figure \ref{fig:Xmas} is owing to the tally for that year being derived from the mid-year numbers.} On average, an object is observed on the order of 10 times per year, during its range of detectability, e.g. while the comet passes through its perihelion. Table \ref{tab:suobs} lists the number of observations from each of the leading five surveys at 5 year intervals back to 2000. The table shows the number of comet observations is relatively small compared with the total observations of small bodies. It also reveals the slowly changing ranks (in order of total observations) in the lead surveys. The output of some very active programs are temporarily diminished (cf. NEAT in the year 2000); each program either upgrades and incorporates more sites, or becomes outpaced by competing surveys, in which case existing survey programs or sites often shift to a priority from discovery to highly productive follow-up.  

Both ground-based and space-based sky surveys have been used to characterize cometary populations. However, the full and systematic utilization of the majority of data obtained by the surveys is in its early stages, with only a handful of instances of the data being used to quantitatively characterize the comet populations.  Much of the initial exploration of these sky survey datasets are centered around characterization of particular comets of interest. \cite{2021RNAAS...5..211D} use Asteroid Terrestrial-impact Last Alert System (ATLAS) data to identify the longevity of 95P/Chiron's 2018 onset of activity. Zwicky Transient Factory (ZTF) survey \citep{2021PSJ.....2..131K}, Transiting Exoplanet Survey Satellite (TESS) spacecraft \citep{2019ApJ...886L..24F}, and NEOWISE \rev{survey}  \citep{2021PSJ.....2...34B} observations were used to monitor and characterize the behavior of 46P/Wirtanen during its 2019 perihelion approach. Investigations of statistically significant samples of comet populations (c.f. \citealt{2021DPS....5330106F}) are likewise facilitated by sky surveys, and are beginning to be analyzed using astroinformatic approaches. \rev{Larger surveys that compile cometary populations to constrain populations statistics are rarer still. \cite{Hicks07} reported magnitudes and Af$\rho$ values (c.f. \citealt{AHearn1984}) for 52 comets observed by the Near Earth Asteroid Telescope (NEAT) between 2001 and 2003, and produced estimates of nucleus size for 25 of the lowest-activity comets in the sample. Searches for cometary activity among asteroids are more common (see also Jewitt et al. in this volume). \cite{2013Wasz} searched for undiscovered main-belt comets, but identified 115 comets in the Palomar Transient Factory (PTF) data taken from 2009 through 2012, listing the maximum and minimum magnitudes observed in the images. \cite{2011Sonnett} observed 924 asteroids, and \cite{2015Hsieh} conducted a large search of main belt objects observed in Panoramic Survey Telescope and Rapid Response System (Pan-STARRS) data to find cometary activity, while \cite{2019Martino} and \cite{2020Mommert} searched for activity amongst asteroids in comet-like orbits.}

\subsection{Survey Biases}
\label{sec:surveybias}

\begin{landscape}
\begin{figure} 
\vspace{-1.5in}
\hspace{-0.9in}
\begin{center}
\includegraphics[width=1.45\textwidth]{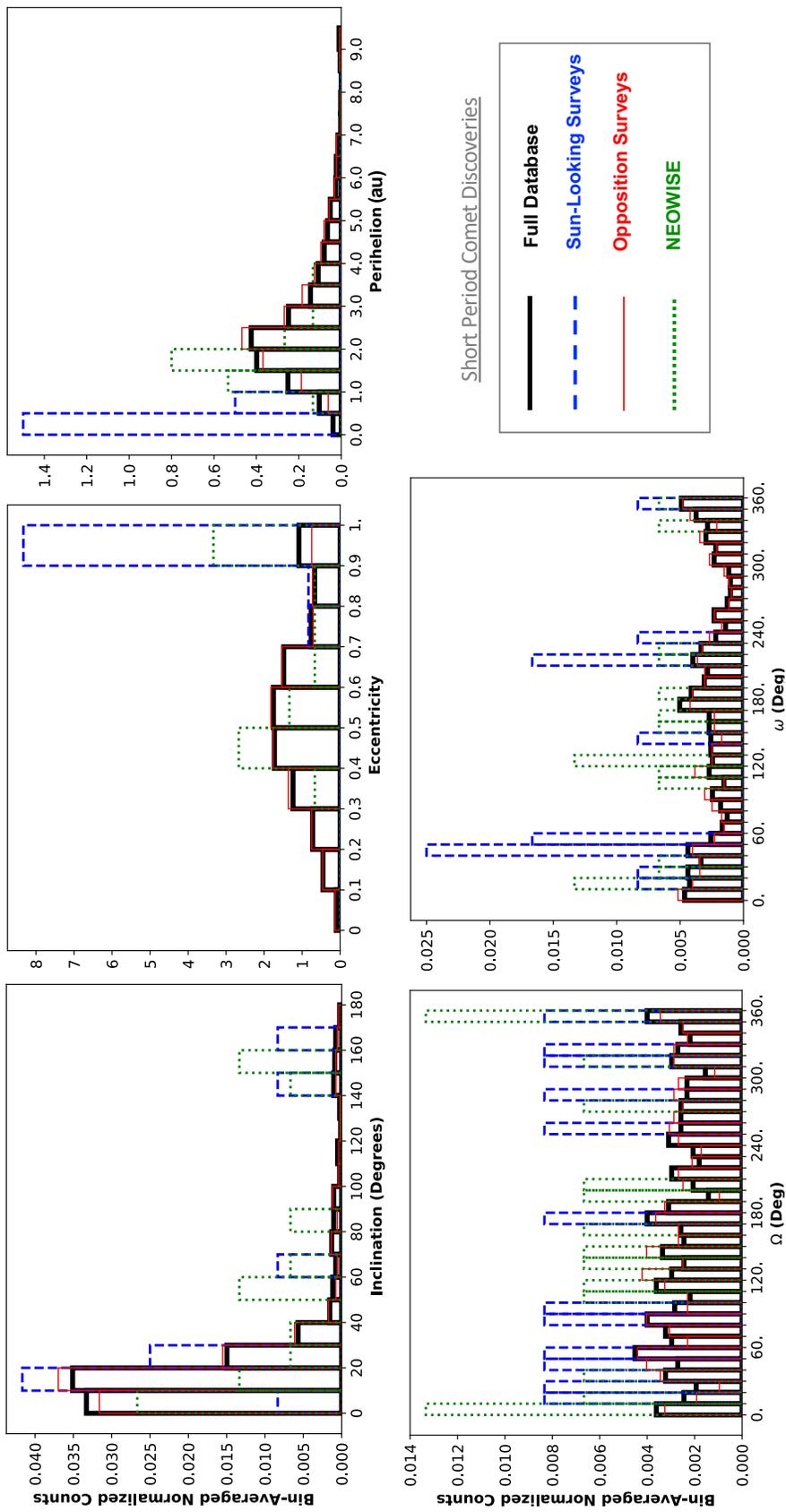}
\vspace{-1.2in}
\caption{The relative distributions of the orbital elements of short period comet (SPC) discoveries from the JPL small bodies database are shown.   The database's SPC populations are shown in total (black histogram), and compared with representative samples of contributions from sun-looking surveys (blue dashed line; SOHO, SWAN, and STEREO comets), opposition pointing surveys (red thin solid line) and terminator pointing (green dotted line; NEOWISE) surveys. Scaled distribution histograms are shown for orbital inclination (upper left), eccentricity (upper middle), perihelion (upper right), ascending node (lower left) and argument of perihelion (middle left). Note that each histogram is plotted with bin-averaged, normalized counts, such that the sum of the counts times the bin-width across all the bins equal 1.}
\label{fig:SPCbias}
\end{center}
\end{figure}
\end{landscape}

\begin{landscape}
\begin{figure}
\vspace{-1.5in}
\hspace{-0.9in}
\begin{center}
\includegraphics[width=1.45\textwidth]{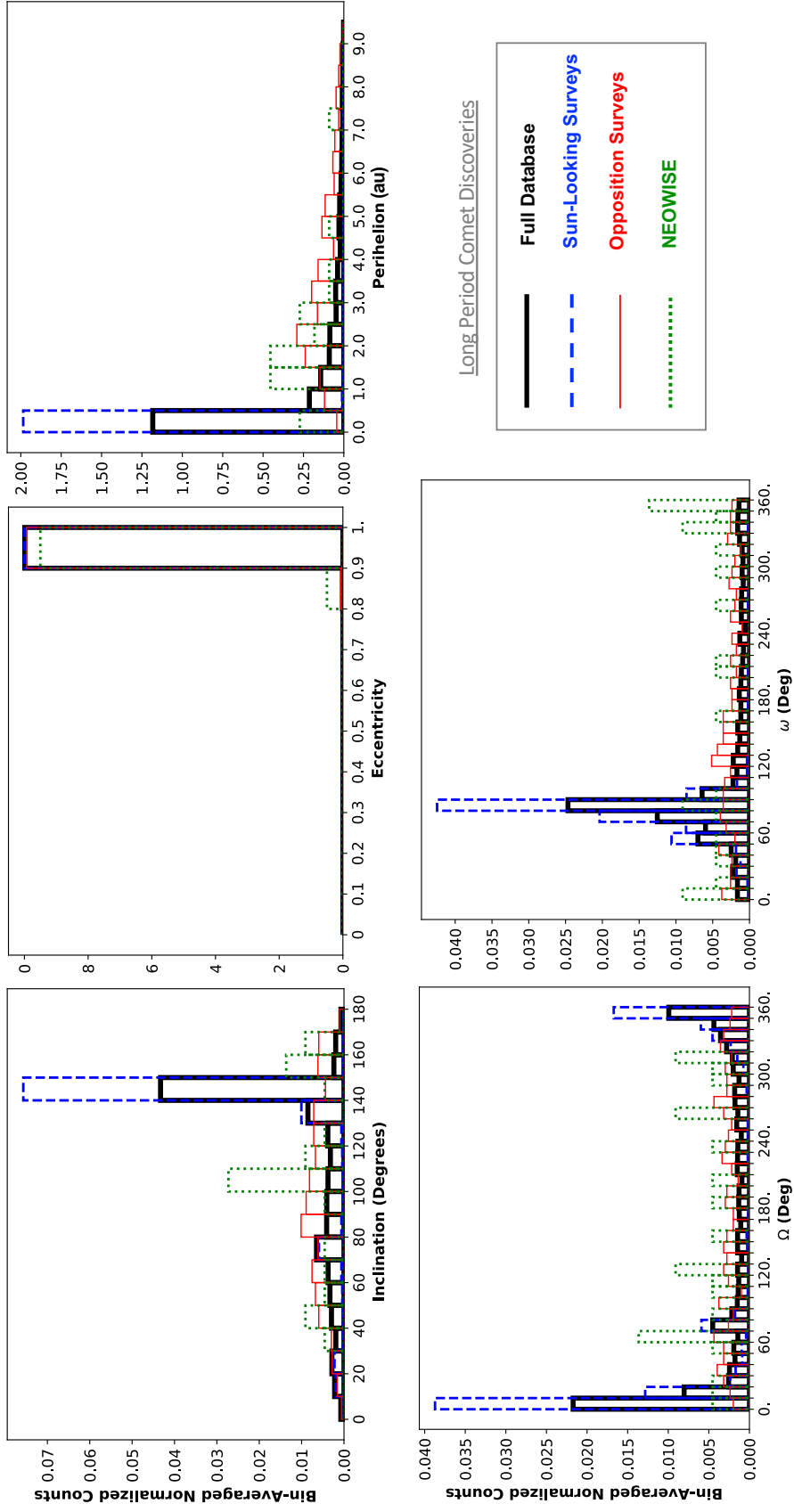}
\vspace{-1.0in}
\caption{Similar to Fig. \ref{fig:SPCbias}, the relative distribution of the orbital elements, here for the long period comet (LPC) discoveries from the JPL small bodies database are shown. The database's LPC populations are shown in total (black histogram), and compared with representative samples of contributions from sun-looking surveys (blue dashed line; SOHO, SWAN, and STEREO comets), opposition pointing surveys (red thin solid line) and terminator pointing (green dotted line; NEOWISE) surveys. Scaled distribution histograms are shown for orbital inclination (upper left), eccentricity (upper middle), perihelion (upper right), ascending node (lower left) and argument of perihelion (middle left).}
\label{fig:LPCbias}
\end{center}
\end{figure}
\end{landscape}

It is abundantly clear from the previous discussion that \rev{sky} and sun-pointing surveys have had a remarkable impact on our statistical understanding of the comet populations. However, these data possess limitations, according to their sensitivities, coverage strategies, and cadences. Non-survey observations remain, therefore, highly valuable to the community, as demonstrated in the discovery of the second interstellar object (cf. \citealt{2021SoSyR..55..124B}). Figs \ref{fig:SPCbias} and \ref{fig:LPCbias} are illustrative of the sample biases that can remain even when short term biases, like those imposed by weather, are removed or averaged over, and how they can be convolved with real population features.  The SPC inclination features are mostly real, and are dominated by the JFC population's clustering near low-inclination orbits. The outliers in inclination, around near-retrograde orbits, have contributions from the active Centaur and Halley-type comet populations. Eccentricity is nearly level, but falls off at near zero values, corresponding to near-circular orbits; the high-eccentricity outliers on short-period orbits are in part strengthened by the near sun-grazing comets seen by sun-looking surveys (SLSs), or higher elongation, or terminator-pointing surveys (TPSs), like NEOWISE.\rev{\footnote{The Near-Earth Object Wide-field Infrared Survey Explorer (NEOWISE) uses the repurposed Wide-field Infrared Survey Explorer (WISE) spacecraft to search for NEOs and other solar system bodies.}} Comets tend to be most active as they approach towards and retreat from their perihelion distance, so the discoveries with the furthest perihelia are made first by the opposition-looking surveys, while the TPSs make the near-earth perihelion discoveries, and the remaining low-perihelion comets are found by SLSs.  The dips near 80 and 270 degrees in the SPC argument of perihelion ($\omega$) distributions roughly correspond to where the tug of Jupiter disrupts SPC orbits. 

For the LPCs, in contrast to the SPCs, the discoveries are dominated by SLSs, and thus a few high-inclination sun-grazing comets, particularly the Kreutz family comets. The NEOWISE weak cluster in inclination, near 105$^{\circ}$, pointed out in \cite{2017AJ....154...53B}, has become mildly more pronounced with additional survey data, and the peaks in ascending node ($\Omega$) and $\omega$ are clustering from the noted Kreutz family comets. In each survey approach, the success with particular populations show statistical outliers that can bias the derived distributions if naively extrapolated from the observed distributions or not carefully removed. 

The large representative samples of comets discovered and observed by sky surveys facilitates analyses of their total populations that lead to constraints on their total numbers. Such derivations are common among other representative populations, for example with NEOs \citep{2012ApJ...752..110M} and Centaurs \citep{1997Icar..127..494J}.  Accurate accounting of factors that affect the survey's detection efficiency, such as observing cadence, pointing pattern and viewing geometry, sensitivity, and weather (for ground-based surveys) are critical to the assessment of the underlying population numbers from the observations.  Owing to these factors being intrinsic to each survey (or instrument/telescope combination), they must be considered for each separate contribution. Comets, however, are different from other populations in that they require an extra layer of accounting to derive the final total population numbers from the observed sample; the brightness variations from activity have to be accounted as well. Comets tend to vary greatly in their brightness throughout their orbit, usually achieving peak brightness around, though often not precisely at, their perihelion. Even surveys with more predictable observing circumstances, e.g. space-based surveys, have an additional significant level of uncertainty on any derived constraints of total populations.

\begin{table}

\caption{Yearly Comet Survey Observations$^{a}$ \label{tab:suobs}}
\begin{tabularx}{\columnwidth}{llll}
\noalign{\vskip 5 pt}
\hline
\hline\noalign{\vskip 5 pt}
 Survey$^{b}$&	Total$^{c}$ &	Comets &	Comet \\
 &Detections$^d$ &Observed& Detections$^d$ \\
\hline
\multicolumn{4}{c}{\bf 2020} \\
\hline
Pan-STARRS  &  12181991	& 344	& 3606 \\
ATLAS	      & 10396137	& 284	& 11787 \\
Catalina    &  10134103	& 335	& 4754 \\
NEOWISE     &  152141  	& 30	& 324  \\
Spacewatch  &	70385   	& 13	& 60   \\
 {\it Yearly Total:}       & 32934757	& 1006	& 20531 \\
  \multicolumn{2}{l}{\it Survey Fraction:}	& {\it 0.18}	& {\it 0.31} \\
 \hline
\multicolumn{4}{c}{\bf 2015} \\
\hline
Pan-STARRS  &  7256500	& 206	& 2533 \\
Catalina    &  3950145	& 176	& 1566 \\
Spacewatch  &	512214   	& 51	& 276   \\
NEOWISE     &  158595  	& 53	& 840  \\
ATLAS	      &  104495  	& 189	& 5708 \\
{\it Yearly Total:}       & 11981950	& 514	& 5484 \\
  \multicolumn{2}{l}{\it Survey Fraction:}	& {\it 0.10}	& {\it 0.09} \\
\hline
\multicolumn{4}{c}{\bf 2010} \\
\hline
Catalina  &  3296494	& 128	& 1119 \\
NEOWISE$^e$    &  2410314	& 111	& 1477 \\
LINEAR  &	 2193193  	& 78	& 1122 \\
Spacewatch    &  852890	& 67	& 462 \\
Pan-STARRS     &  597563  	& 7	& 28  \\
{\it Yearly Total:}       & 9350456	& 391	& 4208 \\
  \multicolumn{2}{l}{\it Survey Fraction:}	& {\it 0.09}	& {\it 0.08} \\
\hline
\multicolumn{4}{c}{\bf 2005} \\
\hline
Catalina    &  2325309	& 132	& 1342 \\
LINEAR     &   2056210	& 98	& 1697 \\
Spacewatch  &	1063276  & 46	& 287   \\
NEAT	      &  549313	& 37	& 207 \\
LONEOS  &  803620	& 66 & 1513 \\
{\it Yearly Total:}       & 6797728	& 379	& 4046 \\
  \multicolumn{2}{l}{\it Survey Fraction:}	& {\it 0.11}	& {\it 0.11} \\
\hline
\multicolumn{4}{c}{\bf 2000} \\
\hline
Spacewatch  &	2149917  & 14	& 134   \\
LINEAR     &   2094140	& 84	& 1682 \\
LONEOS  &  456892	& 50 & 343 \\
Catalina    &  48878	& 7	& 36 \\
NEAT	      &  29	& 2	& 6 \\
{\it Yearly Total:}       & 2814856	& 157	& 2201 \\
\multicolumn{2}{l}{\it Survey Fraction:}	& {\it 0.09}	& {\it 0.11} 
\end{tabularx}
\tablenotetext{a}{Annual totals shown at \rev{five-year} intervals.} \tablenotetext{b}{Non-solar-pointing surveys and follow-up programs.} \tablenotetext{c}{The total includes asteroids and comets.} \tablenotetext{d}{Observations reported to the MPC; more complete \\ summary available at  https://sbnmpc.astro.umd.edu\/.}
\tablenotetext{e}{\cite{2017AJ....154...53B} notes 164 comets observed by WISE/NEOWISE within \\ the year, many retrieved by  stacking. This number represents those \\ detected by the automated detection pipeline. }  
\end{table}

The earliest estimates of background populations based on a modern sky survey was conducted by \cite{2005ApJ...635.1348F}. The author used the Lincoln Near-Earth Asteroid Research (LINEAR) survey to assess the long-period comet population. \cite{2005ApJ...635.1348F} found a total population of $\sim 5 \times 10^{11}$ comets with a nucleus size of roughly a 1km in diameter, roughly a factor of $\sim 2.5$ times that predicted by \cite{1950BAN....11...91O}.  An important detail is that because the small end of the comet size distribution is difficult to measure, many of the population estimates are for lower-bounded size ranges. The difficulty in assessing the comet populations with effective diameters less than a kilometer \rev{often results in values of $\gtrsim 1$ km for the lower bound in size for population comparisons.  The Survey of Ensemble Physical Properties of Cometary Nuclei (SEPPCoN)}  \citep{2013Icar..226.1138F} provided constraints on the JFC population between two and ten thousand objects with diameters of approximately a kilometer or larger. \cite{2017AJ....154...53B}, using the NEOWISE survey data, arrived at a number that fell within the lower end of that range, $\sim 2100$ Jupiter Family comets. Applying a similar technique to the observed LPCs, \cite{2017AJ....154...53B} found a total population of $1.3 \times 10^{12}$ Oort Cloud comets, about twice that of the LINEAR-derived value by \cite{2005ApJ...635.1348F} and also found that the majority of LPCs, about $\sim 60\%$, were already detected by contemporary surveys. It is worth noting that since 2015, the rate of the discovery of non-sun-grazing LPCs has held an average of 6.2 comets per year with perihelia within 1.5 au per year. Most recently, the PanSTARRS survey has been assessed and de-biased to obtain JFC and LPC population constraints \citep{2019Icar..333..252B}. The comparative size constraints will be discuss in Section \ref{sec:nuc} , but population totals for JFC and LPC comets find similar numbers, with $\sim 10^{12}$ Oort Cloud objects as the speculative total.

\begin{table*}
\footnotesize
\caption{Summary of Selected Surveys Since {\sl Comets\ II}\label{tab:summ}}
\begin{tabularx}{\textwidth}{cccccccccl}
\noalign{\vskip 5 pt}
\hline
\noalign{\vskip 5 pt}
\multispan{2}{\hfil $N$ \hfil} & $N_{obs}$ &
$\lambda$ & Location & Designated &
Technique & Telescope & Instrument  & Reference \\
SPCs & LPCs &  &  &       &  Asset & & or Survey & &\\
\hline
\noalign{\vskip 5 pt}
 0 & 150 & 150 & 735, 870 nm & G & A & I & Pan-STARRS 1 & -- &  \citet{2019Icar..333..252B} \\
 95 & 56 & $\sim$3000 & 3.5, 4.6, 11, 22 $\mu$m & S & A & I & NEOWISE & -- &  \citet{2017AJ....154...53B}, \\
 & & & & & & & & & \citet{Bauer2015}, \\ 
 & & & & & & & & & \citet{2015LPI....46.2820K} \\
 100 & 0  & $\sim$200 & 16, 22, 24 $\mu$m & S & A & I & Spitzer & IRS, MIPS &  \citet{2013Icar..226.1138F},\\
 & & & & & & & & & \citet{Kelley2013} \\
 \rev{34} & \rev{0} & \rev{34} & \rev{24 $\mu$m} & \rev{S} & \rev{A} & \rev{I} & \rev{Spitzer} & \rev{MIPS} & \rev{\citet{2007ReachSSTtrails}} \\
 \rev{18} & \rev{4} & \rev{38} & $3.6, 4.5 \mu m$ & S & A & I & Spitzer & IRAC & \citet{2013Icar..226..777R} \\
 \rev{23} & \rev{1} & \rev{30} & \rev{B,V,R,I} & \rev{S} & \rev{A} & \rev{I} & \rev{HST} & \rev{WFPC2} & \rev{\citet{2009Icar..201..674L}} \\
  17 & 44 & 3700 & Ly-$\alpha$ & S & H & I & SOHO & SWAN &  \citet{2019Icar..317..610C}\\
 \rev{44} & \rev{0} & \rev{$\sim$200} & \rev{R} & \rev{G} & \rev{A} & \rev{I} & \rev{various$^b$} & \rev{various$^b$} & \rev{\citet{2011MNRAS.414..458S}$^b$} \\
 0 & \rev{23} & \rev{29} & B,V,R,I & G & A & I & Keck I & LRIS &  \citet{Jewitt2015}\\
 \rev{24} & \rev{6} & \rev{35} & u,g,r,i,z & G & A & I,S & SDSS & CCD &  \citet{Solontoi2012}\\
 \rev{6} & \rev{14} & \rev{25} & \rev{$3-14 \mu m$} & \rev{G} & \rev{P} & \rev{S} & \rev{IRTF} & \rev{BASS} & \rev{\citet{Sitko2004}}\\
  \rev{42} & \rev{0} & \rev{53} & \rev{Johnson R} & \rev{G} & \rev{P,A} & \rev{I} & \rev{Kiso 1.05m} & \rev{CCD} & \rev{\citet{2009IshiguroKasi}} \\
 \rev{100} & \rev{0} & \rev{215} & \rev{various} & \rev{G,S} &\rev{P} & \rev{I} & \rev{various} & \rev{various} & \rev{\cite{MazzottaEpifani2009}}\\
 28 & 22 & 218 & near-UV, visible & G & P & S & UA 1.54m & CCD & \citet{2009Icar..201..311F}\\
 77 & 53 & 558 & near-UV, visible & G & P & S & McDonald & CCD & \citet{2012Icar..218..144C}\\
 11 & 19 & 152 & $ 2.5 - 5 \mu m $ & G & P, A & S & various$^{a}$ & various$^{a}$ & \citet{2016Icar..278..301D} \\
 8 & 12 & 54 & $ 2.5 - 5 \mu m $ & G & A & S & Keck2 & NIRSPEC & \citet{2021AJ....162...74L} \\
 -- & 50 & 152 & Johnson R & G & P, A & I & 0.6/0.9/1.8m Schmidt  & CCD & \citet{2016AJ....152..220S} \\
 $\sim100$ & $\sim100$ & $\sim$1000 & near-UV, visible & G & P & I & Lowell & & \citet{BairSchleicher2021DPS} \\
\hline
\end{tabularx}
\tablenotetext{a}{NASA-IRTF with CSHELL, Keck2 with NIRSPEC, Subaru with IRCS, and the VLT with CRIRES.}
\tablenotetext{b}{A variety of telescopes were used by the same group for the survey. The paper mentioned here compiles all the group's previous results from earlier papers.}
\tablecomments{We limit this table to surveys observing \rev{20 or more} comets.
$N$ refers to the number of comets (of SPCs and LPCs) in the
survey; $N_{obs}$ refers to the total number of observations
made of those $N$ comets; $\lambda$ indicates the primary
wavelength(s) of survey's observation.
In the `Location' column, `G' indicates ground-based, `S' indicates
space-based.
In the `Designated Asset' column, \rev{the original designation, which presumably drove the platform's design requirements}, is listed; `A' indicates Astrophysics,
`P' indicates Planetary, and `H' indicates Heliophysics. 
In the `Technique' column, `I' refers to imaging, and `S' refers to spectroscopy.}
\end{table*}


\newpage

\section{SURVEYS OF COMETARY DUST} 
\label{sec:dust}
\subsection{Broad-band Visible Imaging}\label{sec:dust_imaging}

While spectroscopy in the visible and infrared wavelength ranges is the most diagnostic tool to investigate the composition of cometary dust, including both refractories and ice compounds, photometry, especially in the visible wavelength range, has been the most used technique to characterize a large number of objects. Imaging with broad-band filters in the visible wavelength range, yielding color indices, enables measurement of the solar light scattered by cometary dust particles, from which it is possible to infer first order compositional information such as particle size and ice to refractory ratios. Analysis of possible correlations between optical colors and orbital parameters can highlight composition diversity among sub-populations possibly attributable to variations at formation in the protoplanetary disk and/or evolutionary processes.

\begin{figure*}[ht!]
\begin{center}
\includegraphics[width=\linewidth]{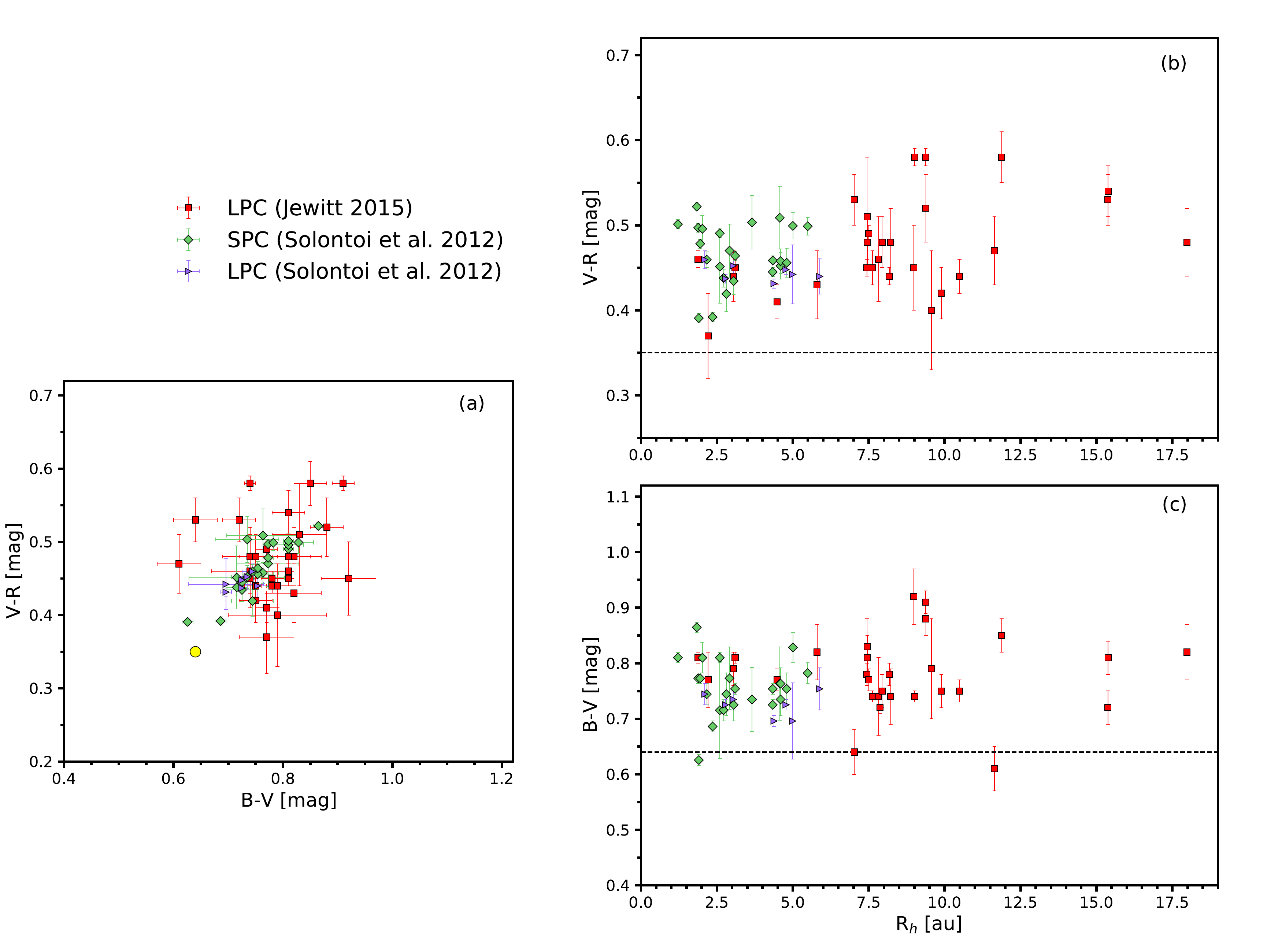}
\caption{(a) Color–color diagram comparing LPCs from \citet{Jewitt2015} with LPCs and SPCs from the SLOAN survey reported by \citet{Solontoi2012}. The color of the Sun is
marked by a yellow circle. In (b) and (c) the $V-R$ and $B-V$ colors of the same comets, respectively, is shown as a function of heliocentric distance in au. The dashed horizontal line represents the color of the Sun.}
\label{fig:Figure1_sp}
\end{center}
\end{figure*}

\citet{Solontoi2012} compiled $u$, $g$, $r$, $i$, $z$ band photometry of 26 active comets (6 LPCs and 20 JFCs) observed by the Sloan
Digital Sky Survey \citep[SDSS,][]{York2000} spanning a range of heliocentric distances between $\sim$1 and $\sim$6~au \rev{(observations of unresolved comets are not considered)}. \citet{Jewitt2015} extended the work by \citet{Solontoi2012} by presenting new $B$, $V$, $R$ photometric measurements  for \rev{23} active LPCs obtained with the 10 m diameter Keck I telescope at
Mauna Kea and the Low Resolution Imaging
Spectrometer (LRIS) camera \citep{Oke1995}. This data set not only quadruplicated the number of LPCs for which colors are available, but also broadened the heliocentric distance range with measurements up to 18~au from the Sun. After transforming the measurements by \citet{Solontoi2012} in the Sloan filter system into $BVRI$ photometry \citep{Ivezic2007}, \citet{Jewitt2015} investigated possible correlations between optical colors
and orbital parameters for the combined SDSS+Keck data set (see Figure~\ref{fig:Figure1_sp}). No significant difference was found between the mean colors of active SPCs and LPCs (Figure~\ref{fig:Figure1_sp}, panel a), suggesting the lack of compositional variation between these two groups. The author pointed out the agreement between this finding and gas-phase studies reported by \citet{AHearn2012} and \citet{Cochran2015} and attributed it to the idea, already put forth by \citet{AHearn2012}, that JFCs and LPCs formed in largely overlapping regions of the protoplanetary disk.

\rev{No trend was found between the $B-V$ and $V-R$ colors and heliocentric distance ($R_{h}$, Figure~\ref{fig:Figure1_sp}, panels b and c). \citet{Jewitt2015} attributed this evidence to 1) ice-to-dust ratio being on the order of only a few percent and 2) small particles not being abundant enough to dominate the scattering cross section. The first conclusion relies on the idea that solid state water has been detected in cometary comae \citep{Davies1997,Kawakita2004,Yang2014,Protopapa2018}, it is stable at large heliocentric distances with sublimation rates varying inversely with heliocentric distance and it is bluer than refractory materials. Therefore, an ice-to-dust ratio larger than few percent would lead to bluer colors with increasing heliocentric distance, contrary to what was observed. The second conclusion leans upon the expectation that, at large heliocentric distances, given the lower gas flow, the mean size of the ejected particles should fall into the Raleigh regime ($X=\pi D/\lambda <<1$, where $D$ is the particle diameter and $\lambda$ is the wavelength of observation), yielding to blue colors \citep{Bohren1983}. However, \citet{Gundlach2015} found, through numerical modeling and laboratory results, that the size range of the dust aggregates able to escape from
a comet nucleus into space widens when the comet approaches
the Sun and narrows with increasing heliocentric distance. This is because the tensile
strength of the dust aggregates decreases with increasing
aggregate size. Therefore, at large heliocentric distances, given the lower gas flow, only large aggregates would be lifted off the nucleus. These arguments, which rely on the
assumption that comets have formed by gravitational instability \citep{Skorov2012,Blum2014}, weaken the conclusion that small particles are not abundant enough to dominate the scattering cross section.}

Not all broad-band comet surveys provide color information, but still measure activity and dust production. \cite{2016AJ....152..220S} measured A$f\rho$ values and the coma slope parameters for 50 LPCs with known activity beyond 5 au. \rev{The $A f \rho$ quantity is the product of albedo ($A$), filling factor of the grains within the field of view ($f$), and the linear radius of the field of view at the comet ($\rho$), while the slope parameter is defined as $d$ log A$f\rho$ / $d$ log $\rho$ \citep{AHearn1984}. These parameters are diagnostic of the comet activity and are a proxy for the dust production rate and the morphological appearance of the coma.} \rev{ \citet{2016AJ....152..220S}} divided the LPC sample into \textcolor{black}{dynamically new ($a > 10^{4}$ au)} and returning comets, and found that on average the A$f\rho$ of \rev{dynamically new comets} significantly exceed those of the recurrent LPCs, similar to the earlier work of \cite{2009Icar..201..719M} with a smaller sample. Furthermore they found that \rev{new comets} usually exhibit negative (shallow) slope ($d$ log A$f\rho$ / $d$ log $\rho$) parameters, and symmetric comae. The comets which were strongly active beyond 10 au, they found, tend \rev{to} have a smaller increase in A$f\rho$ with decreasing heliocentric distance.
\subsection{Thermal dust}
Analogously to the $A f \rho$ parameter introduced by \rev{\citet[][see Section \ref{sec:dust_imaging}]{AHearn1984}} as a proxy for the dust production rate measured at visible wavelengths for scattered light observations, \citet{Kelley2013} \rev{used} the $\epsilon f \rho$ parameter for observations of thermal emission from comet comae \rev{introduced by \cite{Lisse2002}}. \rev{The effective emissivity of the grains is parametrized by $\epsilon$, while $f$ is the areal filling factor within an observed aperture of radius $\rho$.} \citet{Bauer2015} carried out a comparison between the log $\epsilon f \rho$ as derived from the WISE W3 (12~$\mu$m) and W4 (22~$\mu$m) channel dust emission and the log $A f \rho$ values  derived from the W1 (3.4 $\mu$m) channel assuming dust signal was dominated by reflectance. For the comets observed at heliocentric distances exceeding 3~au, the difference between the values ($\sim 0.86 \pm 0.1)$ were found to be consistent with dark dust with emissivity near 0.9. Comets observed at heliocentric distances inside 3~au deviated from this trend, possibly owing to \rev{the different size range of aggregates lifted by activity at smaller rather than larger heliocentric distances,}  given the higher
gas flow, or to the more pronounced thermal emission component within the 3.4 $\mu$m band signal for the dust at \rev{ small} distances. Outside of 3~au, the sample was dominated by LPCs; only 4 SPCs were measured outside of 3~au. 

\begin{figure}[ht!]
\begin{center}
\includegraphics[width=\linewidth]{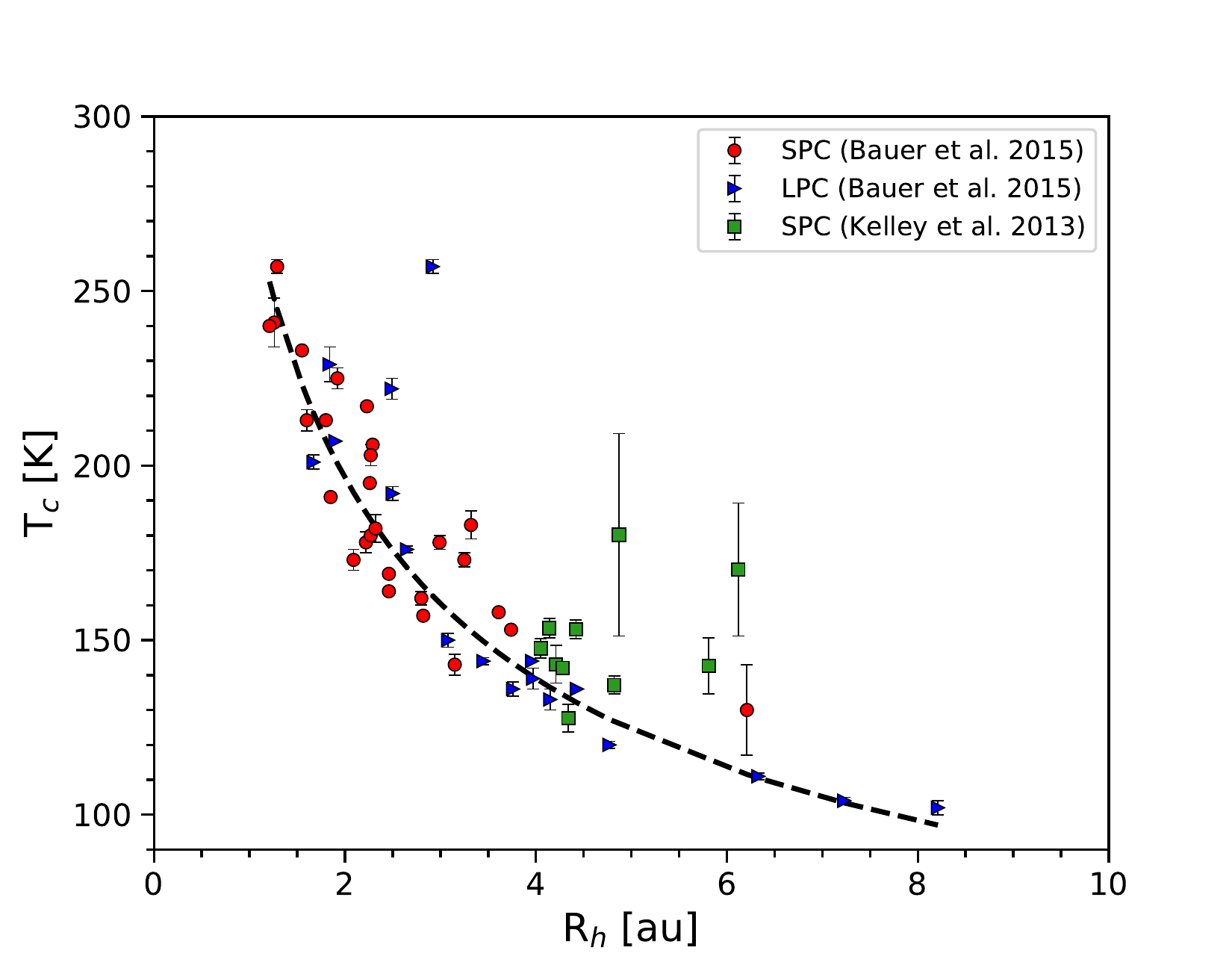}
\caption{Coma color temperatures as a function of heliocentric distance comparing LPCs from \citet{Bauer2015} with SPCs reported by \citet{Bauer2015} and \citet[][only color temperatures of the dust centered on cometary nuclei are shown]{Kelley2013}. The dashed line represents the temperature of an isothermal blackbody ($278 R_{h}^{-1/2}$).}
\label{fig:Figure2_sp}
\end{center}
\end{figure}

The physical temperature of cometary dust grains is a function of composition, grain size and morphology. An approximation for the true physical temperature of the grains is given by the color temperature $T_{c}$ determined through analysis of the thermal spectral energy distribution over a defined wavelength range \citep[e.g.,][]{Wooden2002,Kolokolova2004}. \rev{The temperature excess of a cometary coma over that of a blackbody is usually parameterized by the superheat parameter $S_{heat}$ introduced by \citet{Gehrz1992} and defined as the ratio of the color temperature and the temperature of an isothermal blackbody
sphere in LTE at the same heliocentric distance ($T_{BB} = 278 R_{h}^{-1/2}$).} \citet{Bauer2015} reported color dust temperatures based on the WISE W3 and W4 band thermal fluxes for 24 SPCs and 14 LPCs while 
\citet{Kelley2013} reported color temperatures for 15 SPCs through analysis of the 16- and 22-$\mu$m \textit{Spitzer Space Telescope} MIPS observations \rev{(Fig.~\ref{fig:Figure2_sp})}. No differences were found between SPCs and LPCs \citep{Bauer2015}. \rev{Overall, cometary comae color temperatures measured with WISE displayed a slight excess of 1.6 $\pm$ 0.1\% (the error-weighted mean of all the superheat measurements is $\bigl\langle T_{c}/T_{BB}\bigr \rangle = 1.016 \pm 0.001$) and were found to be consistent with isothermal bodies with emissivity $\sim$0.9 and albedo 0.1 at 3.4~$\mu$m \citep{Bauer2015}. A more significant temperature excess was found by \cite{Kelley2013}, who reported an error-weighted mean of all their color-temperature measurements of $\bigl\langle T_{c}/T_{BB}\bigr \rangle = 1.074 \pm 0.006$. This translates in the color temperature of the dust to be 7.4 $\pm$ 0.6\% warmer on average
than an isothermal blackbody sphere in LTE. An important caveat to consider when assessing the validity of color temperatures obtained from broad-band infrared photometry is the possible presence of emission features, specifically silicate features, above the continuum which could affect the thermal flux measurements and consequently, the color temperature estimates. Therefore, to properly characterize the thermal properties of cometary grains, spectroscopic data of a large number of comets are required. \cite{Sitko2004} analyzed spectroscopic data over the wavelength range 3-14~$\mu$m of 20 comets belonging to different dynamical classes and found cometary grains radiating at temperatures in excess of that of a blackbody at
the equilibrium temperature expected for their heliocentric
distances. This effect is expected for a grain
population that includes a significant fraction of the grains
with sizes smaller than the wavelength of light being radiated,
in this case from 3 to 14~$\mu$m. Additionally, \cite{Sitko2004} found a strong correlation between excess temperature and silicate band strength for dynamically new and long-period comets, confirming the results by \citet{Gehrz1992} and \cite{Williams1997}. The majority of
Jupiter family objects were found to deviate from this relation. To explain the different trend between JFCs and dynamically new comets, the authors put forth the idea of a radial gradient in the
size distribution of silicate grains within the protoplanetary disk. Further observations are required to confirm this finding.}

\rev{Mid-IR broadband images of comets were found to be well suited not only for characterizing the properties of the dust grains but also to investigate the activity level of comets. \cite{Kelley2013} using \textit{Spitzer Space Telescope} images acquired as part of the SEPPCoN survey, investigated the activity of 89 JFCs at 3–7 AU from the Sun and found that activity, detected in at least $\approx 24\%$ of the comet sample, is
significantly biased to post-perihelion epochs. Additionally, \cite{Kelley2013} suggested a bias in the discovered JFC population given that low-activity comets with
large perihelion distances were found missing from the survey sample. The link between activity level and present and historical orbital parameters was also investigated by \cite{MazzottaEpifani2009} through analysis of a sample of 90 SPCs as seen from the ground and space at
heliocentric distances greater than 3~au. This analysis led to several findings including, but not limited to, the higher likelihood of SPCs being active post-perihelion rather than pre-perihelion, similarly to what was found by \cite{Kelley2013}, the lack of a sharp cut-off in heliocentric distance marking the activity-fade, and a weak trend of comets with increasing perihelion
distance to be more likely active at large heliocentric distance.}

Not all analyses of thermal dust is based on flux measures. \rev{\cite{2007ReachSSTtrails} conducted a survey of 34 JFCs at 24 $\mu$m using the Spitzer Space Telescope's Multiband Imaging Photometer for Spitzer (MIPS), and found that the majority (27) of the comets exhibited detectable trails. \cite{2009IshiguroTrails} confirmed the prevalence of JFC-associated dust trails, and found that 6 out of 42 Jupiter Family comets exhibited trails detectable at visual-band wavelengths.   By over-plotting zero-velocity syndynes (cf. \citealt{1968finson}), \cite{2007ReachSSTtrails} found that the dust particles were dominated by mm and cm-sized dust, and that the size distribution of the dust particles is not accurately modeled by a single power-law.} \cite{2017ApJ...838...58K} and \cite{EmilyKphdthesis} demonstrated a novel technique of fitting the cometary dust in the thermal infrared observations using WISE/NEOWISE data. Subsequent application of the technique \citep{2015LPI....46.2820K} to a sample of 89 comets shows that such techniques can \rev{elucidate} the behavior of dust output of comets when they are most active. Using radiation pressure to sift the different ranges of particle sizes 
in the tail in combination with the dust thermal signal, they found that much of the 
dust is emitted preceding perihelion, and that the dust mass is primarily residing in millimeter to centimeter sized grains.

\section{SURVEYS OF COMETARY GAS} 
\label{sec:gas}

By nature cometary gas surveys are slow, requiring multi-year commitments to observing the same sorts of objects as they become available one at a time. Whether waiting for a comet bright enough, close enough, or the right phase angle, there may only be one or two comets available per year for which a  detailed study of multiple species is feasible.  Spacecraft visits may show us a snapshot, but the long term surveys of composition have been critical to telling us what comets have in common, when a snapshot observation is surprising or different, and are beginning to help us understand how comets evolve. Surveys of cometary gas have been critical to the development of our understanding of what comets are, what they are made of and how that composition relates to the origin of our Solar System.  The history of how surveys of cometary gas have shaped our understanding of comets is already well summarized by \citet{Cochran2015}, \citet{2015SSRv..197...47B}, \rev{as  well as {\it Biver et al.} (this work)} so here we shall only note a few key points and recent developments.

\subsection{Ground-based Surveys}
\label{sec:gbgobs}

Early optical spectroscopic surveys are what told us that while all comets contained the same molecular species, that those species came in differing abundances. This led to what has become the grail of modern cometary gas studies: whether there are compositional classes of comets that can be tied to dynamical origin, and thus constrain our knowledge of the composition and temperatures in the protoplanetary disk.  \rev{The promising early results of \citet{AHearn1995} demonstrated that there exits class of comets, dominated by JFCs, that are depleted in carbon-chain molecules. They were followed by multiple groups confirming this result, though finding enough OCCs in the carbon-depleted class that the tie between depletion and dynamical origin is clearly not a simple one (\citealt{2012Icar..218..144C}; \citealt{2009Icar..201..311F}; \citealt{2011Icar..213..280L}; \citealt{BairSchleicher2021DPS}). } Infrared and millimiter wave surveys have made the picture more complicated with no consensus on a classification scheme (c.f. the compilations of \cite{2011ARA&A..49..471M},   \citealt{Bockelee2017RSPTA.37560252B}, and 
\citealt{2016Icar..278..301D}), but new analyses of some of these observations such as that of \citet{2021AJ....162...74L} are beginning to find ways to disentangle the numerous abundances and find patterns.

\cite{2021AJ....162...74L} also demonstrate the necessity of continued surveys in the IR and sub-millimeter wavelength regimes. The authors admit significant limitations to the  20 comets within their survey (or the 33 in \citealt{2016Icar..278..301D}). \cite{HarringtonPinto2021DPS....5321005H} is working to compile matched observation of comets with CO and CO$_2$ production rate constraints. Yet, \cite{2021AJ....162...74L} affirm that presently, the overlapping samples across optical through the sub-mm are too sparse to deconvolve compositional states from cometary origin and evolutionary effects. In order to successfully make sense of the compositional trends, the species detectable over the full range of wavelengths must be characterized for numbers of comets comparable to those where optical spectroscopic analyses are presently available.  



    While we generally think of gas surveys as a means to understand relative composition, the physical state of the gas species is a key detail in understanding comet comae as well.  Outflow velocity is a critical characteristic in determining gas production rates, dissociation scales, and total mass loss. \rev{Line-of-sight velocity of the gas can be measured via spectroscopy done at sufficient resolution to detect doppler line broadening of emission lines. Since coma gas velocity is generally close to $\sim 1$ km/s, sufficient velocity resolution is currently achievable with ground based observing via radio observations at submillimeter and longer wavelengths.  Furthermore, when radio spectroscopy includes spatial mapping, and observations are made at a 90 degree phase angle the survey can distinguish potential asymmetries in sunward and anti-sunward outgassing velocities.}  In their survey mapping the OH coma at 18 cm in 28 comets \cite{LovellHowell2015DPS....4741514L} found the gas outflow velocity beyond 1 au was 0.8 km/s regardless of size, production rate or direction.  While at this time \rev{the Atacama Large Millimeter/submillimeter Array (ALMA)} is still too new for there to be enough observations of comets to qualify as a survey: it's potential to spatially map both coma compositions and velocities will produce a unique survey to look forward to.

\subsection{Spacecraft Surveys}
\label{sec:sccgobs}
The advantages of spacecraft are clear, and partially addressed in section\ref{sec:intro}, but for the purposes of studying gas production have additional advantages. The often higher resolution provided by space platforms can facilitate the measurement of product decay scales and associations with nucleus orientations and features (c.f. \citealt{2016MNRAS.462S.156F}), However, space-based telescopes and instruments are often designed with both solar system and non solar system targets in mind. Hence, band-passes, resolution, and spatial scales may be only moderately suited to the measurements used to place constraints on the given species. \\

The Solar Wind ANisotropies (SWAN) instrument on the SOHO spacecraft, for example, was designed to measure H-alpha line emission associated with large scale structures in the solar wind \citep{1995SoPh..162..403B}. However, comets manifest h-alpha emission, $90\%$ of which is produced by water dissociation mechanisms. Hence, SOHO's SWAN instrument has been effectively used to measure pre- and post-perihelion production of water in 61 comets \citep{2019Icar..317..610C}. These have provided power-law relationships for water production ($Q_{H2O}$) vs. heliocentric distance ($R_h$) in 44 LP and 17 SP comets. \cite{2011AJ....141..128C} demonstrated the methodology employed in the analysis. Power law fits were provided for the pre and post perihelion approaches of the comets using the relation: 
\begin{equation}
Q_{H2O} = Q_{_{1AU}} {R_h}^p
\end{equation}

where Q$_{1AU}$ is the water production rate extrapolated to when the comet is at $R_h  = 1AU$, and p is the "slope" parameter (manifested as a slope in log units). Comparisons were made of the water production slopes according to dynamical sub-classes of LPCs, \rev{dynamically new} OCCs (with semi-major axis values greater than 20000AU), and SPCs, and compositional sub-classes of "carbon-depleted" and "typical" LPCs and SPCs \citep{AHearn1995}. The \rev{dynamically new} OCCs or younger comets tended to have less variation in slope values, clustering around values of $\sim -2 \pm 1$, and exhibited a possible steepening in slope as LPCs dynamically aged. For short-period comets with measured nucleus effective radius values, larger fractional active area correlated with comets with larger perihelion distances, consistent with less processing. The correlations with compositional classifications, and LP pre- and post-perihelion observations, were inconclusive. \\

\rev{Spacecraft observations can provide unique opportunities to assess gas species such as CO$_2$ which are not available to ground-based observations.  CO$_2$ and CO are two drivers of cometary activity,} outpacing water production in a limited range of circumstances, and their out-gassing in these cases provides the dominant means of ejecting dust from the nucleus into the coma. CO can be detected from the ground at sub-millimeter and infrared wavelengths, but this is not the case for CO$_2$ emission, which is blocked by the absorption of CO$_2$ present in earth's own atmosphere. Alternative means of detecting CO$_2$ are under investigation (e.g. \citealt{2013A&A...555A..34D}, \citealt{2016Icar..266..249M} and \citealt{2019AJ....158..128M}) but direct detection, usually from the 4.26 $\mu$m infrared emission line, is the current means of assessing production rates. Both Spitzer Space Telescope's Infrared Camera's (IRAC's) 4.5 $\mu$m imaging band and the WISE/NEOWISE $4.6 \mu$m band contain both the infrared CO$_2$ and CO ($4.67 \mu$m) emission feature. The CO band relative to the CO$_2$ band is on the order of 11 times weaker. However, without an accompanying CO observation from ground-based assets or spectroscopic data, there is no definitive way to determine which species causes excess in the $\sim 4.5$ micron channels of these two spacecraft. Furthermore, dust thermal emission signal dominates over CO+CO$_2$ excess at heliocentric distances within $\sim 2$ au. For comets at smaller heliocentric distances, the dust thermal signal must be well-characterized or the emission excess extremely pronounced (or both) for successful detections.  
 That being said, \cite{2013Icar..226..777R} produced measurements of 23 comets with CO or CO$_2$ emission using SST, and \cite{Bauer2015} measured 39 comets with CO or CO$_2$ excess using the WISE/NEOWISE (hereafter NEOWISE) survey data. \cite{2013Icar..226..777R} attributed the majority of the excess to CO$_2$ production. Additionally, by comparing with literature measurements of $Q_{H_{2}O}$ the SST survey concluded that water sublimation remained dominant out to $\sim 2.8$ au for most comets, and the highest resolution IRAC images suggested more localized active regions for CO$_2$ production. The NEOWISE results suggested that approximately a quarter of comets observed had significant CO or CO$_2$  excess, and a Q$_{CO_2}$ proportionality $\sim {R{_h}}^{-2}$ within 4 au.  Outside 4 au LPCs tended to be the producers of CO or CO$_2$, and \cite{2021PSJ.....2...34B} attributed that to LPCs possibly being more CO-rich. Recent work by \cite{2021DPS....5321017G} extends these analyses with an additional 52 comets observed by NEOWISE in 2014.  It is also important to note that both the SST and NEOWISE surveys referenced the \cite{2012ApJ...752...15O} Akari spacecraft results. Though, with a sample of 18 comets, smaller than the survey number threshold considered here, the attribution of CO$_2$ as the main species with the SST and NEOWISE surveys was at least in part based on these measurements. Furthermore, the Akari spacecraft spectra provided simultaneous water production comparisons and demonstrated that LPCs were generally producing CO$_2$ at greater distances and in some instances outpacing water production. About a fifth of the CO$_2$ producers had CO production that outpaced CO$_2$ production rates for the Akari-observed sample.

\section{SURVEYS OF COMETARY NUCLEI} 
\label{sec:nuc}


The basic parameters of nucleus observations have been described
by (e.g.) \citet{1991ASSL..167...19J}, but 
surveys of nuclei have historically been difficult to perform
due to the problem of coma confusion. While we as a field
have continued to make good progress on observing nuclei
since the time of
the predecessor volume, we discuss here some of the effects
that can fool us into misinterpreting nucleus photometry and
thereby lead to systematic inaccuracies. 

For a thorough
review of the current understanding of nucleus ensemble
properties, we refer the reader to Knight et al. (this volume),
who discuss what we currently know about the sizes, shapes,
spin states, scattering properties, and thermal properties
of nuclei. One overarching result that is clear from 
such a compilation is that a survey that samples the full
diversity of variation, and that samples enough nuclei to
drive down the Poisson noise in each sampling bin, is crucial
for being able to take the next step in interpreting the
distribution in the context of origins and evolutionary
processes.

The most fundamental (and ongoing) problem is perhaps that of separating
the coma's flux from the nucleus's flux.
In many cases the comet has an extended coma
within which a point-source is embedded, so it is obvious
that the comet is showing us not just light from the nucleus.
A significant step forward in handling these cases came with
the development of empirical coma-fitting routines that could
photometrically separate the contributions from nucleus and coma
\citep{1996Icar..119..370L, 2004come.book..223L}, and such techniques have 
been used in several nucleus surveys
\citep[e.g.][]{2013Icar..226.1138F,2017AJ....154...53B}.
The technique has proven to be successful as evinced by its success
at finding the size of nuclei that were then observed directly
with resolved imaging by visiting spacecraft
\citep[e.g.]{1998A&A...337..945L,2009PASP..121..968L}.
The limitation of this technique arises when the contrast between
the nucleus and coma is too low, i.e. if a large fraction
of the light in the central pixels
is from the coma
\rev{\citep{2018PASP..130j4501H}}. 
This can happen if the coma is particularly strong
or if the comet is distant. 
\rev{While comets are still most often imaged
at visible wavelengths, imaging in the infrared, if the coma
is still sufficiently well-detected, is generally more likely
to result in a robust extraction
\citep{2020DPS....5231604B}. For one thing, the most 
optically-active grains at some infrared wavelengths
usually provide less total surface area (for
a typical size distribution) than those in the visible. 
Also, at thermal wavelengths, the 
nucleus will generally be hotter than the surrounding dust 
grains (since an area on the nucleus only emits into $2\pi$ sr vs.
the $4\pi$ sr dust grains emit into). Both of these effects 
would tend to increase the nucleus-coma contrast.}
The other scenario that can be problematic 
depends on grain outflow dynamics; 
if the coma's surface brightness profile at a given
azimuth deviates
significantly in the inner coma from what is measured in the outer coma,
the extrapolation can yield incorrect results. An additional, but related, 
aspect to this is that the method assumes that the light
from the extracted point-source is all from the nucleus, which may not be 
true. \rev{One} example of 
a difficult case is that of C/1995 O1 (Hale-Bopp),
where the extreme dust production  
complicated the extraction
of the nucleus's signal \citep{2002EM&P...89....3F}. 

There is also the case of a comet that appears as a point-source -- and
hence one might assume that the comet is inactive --
but yet the photometry indicates that there is excess light. The
prototypical comet for this situation is comet 2P/Encke 
\citep{2001DPS....33.2006M,2005Icar..175..194F}. Generally the more distant
the comet, where the linear width at the comet
of the (angular) point-spread function is larger,
the easier it is to hide a dust coma within the seeing disk.
However it seems that this phenomenon alone cannot explain the
specific situation with 2P/Encke;
\citet{BPMMSTh} showed that for a particular dataset (and the
observing conditions that went with it) where the comet 
looked entirely point-source like, only about $\sim$20\%
of the flux could be from a steady-state $1/\rho$ dust
coma. Any more than that would be revealed as wings in the comet's
profile. The analysis did not assess other coma shapes, so it is possible
that a large-grain coma, with particles moving below
escape velocity around 2P/Encke's nucleus, could play a role.

Even when one is sure that the light from a comet is all or
nearly all from the nucleus, the interpretation can still
be muddled if the observation is just a snapshot. The rotational
context is often necessary to be sure of what one is actually
measuring. Fortunately, this problem is not as terrible as it
may seem as first. E.g., \citet{2004come.book..223L} show that 
most measurements even without rotational context will still
often be within $\sim$90\% of the correct answer anyway. Furthermore, for survey data such rotational variations in profile often may be averaged over, as data may span days.

Ideally temporal coverage will extend all around the orbit
so as to understand not only the spin period but also the spin
axis direction and some shape information. 
That often presents a problem, since one may only
be able to see one region of a nucleus from Earth when the comet
is highly active. Of course we now also have much observational
evidence that comets change their spin states on orbital
(and shorter!) timescales as well
\citep[e.g.][]{},
so measurements obtained at multiple epochs may be 
challenging to fold
into each other in the classical ways \citep[e.g.][]{1978ApJ...224..953S}.

\subsection{Sizes}
\label{sec:sizes}

Assessing the size distribution of cometary nuclei 
by definition requires an extensive survey, since it can only be
measured by having a sufficient number of targets. 
One significant problem is that there is always a
diameter above which the sample is sufficiently representative
for a robust analysis. Though that size lower-limit is dependent on the survey's sensitivity, it is not often clear where that
critical diameter is (see also \ref{sec:surveybias}). A plot of the cumulative size
distribution (CSD) of nuclei $N(>D)$ (where $N$ is the number of
nuclei with diameter larger than $D$) always
shows a flattening at small ($\sim$1-2 km) diameters;
equivalently, a plot of the size-frequency distribution (SFD)
$n(D)\ dD$ (where $n$ is the number of nuclei with diameter
between $D$ and $D+dD$) shows a dropoff toward zero at small diameters.
This is at least partially due to the fact that (a) our
discovery of such comets is less efficient to begin with,
and (b) those are often the very comets for which it is hardest
to determine accurate nucleus information because the coma
obscures the nucleus signal and/or the nucleus is just too faint.
\rev{In any case, clearly such a feature of the CSD or
SFD adds difficulty when trying to fit, say, a power law.}

A more robust solution is to 
assess what the observational biases are in the discovery of the
comets. These observational biases will often lead to one or two \rev{effects} that can accounted for in the \rev{CSD or SFD} itself and so will yield a more realistic distribution. This is a challenging task \rev{however}
when comet discoveries are made by a wide range of facilities.
For the smallest comets ($\sim$1 km diameter and below) it is especially
difficult. 

For the JFCs, ideally a survey would either sample a significant fraction 
of the known comets or make a thorough sweep of the sky to discover
the population, including observing the known comets. There must be sufficiently
robust software to identify an object as being active. 
At time of writing there are \rev{over} 600 JFCs 
known\footnote{See \url{https://physics.ucf.edu/~yfernandez/cometlist.html}.}.

For the LPCs, the additional problem, in contrast to the JFCs, is that
the comets are simply not visible for as long a period of time. A JFC
will return again and again to provide (at least theoretically) multiple
observing chances over several decades. 
An LPC is viewable throughout its perihelion only once over the lifetime of the survey.  The LPCs are often more active than the JFCs as well, making it harder to extract nucleus properties. 

Given the number of complications that come with studying
nuclei, it may not be surprising that there is some divergence
in results regarding the size distribution, and 
different methodologies 
complicate the picture. For example, 
the LPC nucleus distribution reported
by \citet{2017AJ....154...53B} \rev{comes} 
from NEOWISE observations in the
infrared \rev{and makes use of} the coma-extraction technique. 
\rev{Additionally, there is the LPC distribution reported by
\citet{2019Icar..333..252B}, which instead comes from 
visible-wavelength 
imaging via the `nuclear' absolute magnitudes in JPL's 
database\footnote{See \url{https://ssd.jpl.nasa.gov/tools/sbdb_query.html}.}, and a model of activity to correct
those absolute magnitudes -- which generally are not 
representative of the nucleus alone -- for coma contribution.
Thus direct comparisons between the two studies could be difficult.
Independent estimates of the particular LPCs in the
two studies would be a useful check,
but such estimates are sparse. It can be noted that
a comparison of JFC (not LPC) nucleus diameters in
the \citet{2017AJ....154...53B} work with those in
the SEPPCoN survey \citep{2013Icar..226.1138F} and
with those from spacecraft encounters show
a reasonable match to within 25 percent.
}


Another potentially useful approach to get around the problem
of contaminating coma is to restrict a survey to objects
that are known to be inactive or only very weakly active. 
For example, as part of the
overall ExploreNEO survey, \citet{2015AJ....150..106M} report
observations of several dormant or extinct comets.
However the connection between the size distribution of such highly-evolved
comets and of the active JFCs is still to be determined. Extinct comets
are, by definition, 
after all the survivors of an active lifetime that for many
comets includes significant fragmentation (and thus potentially
a large change in size) if not total disintegration.

%
%

\section{FUTURE SURVEYS OF COMETS} 
\label{sec:fut}

Surveys will continue to be vital for us to probe
the ensemble properties of comets and specifically to understand
the full diversity of the population. While flybys and rendezvous
of specific comets will of course provide detailed studies of such 
objects and phenomenological first-hand accounts, it is important that the comet community
continue to take advantage of ground-based and space-based
telescopic assets that can shed light on a representative cometary
sample. This includes making use of facilities whose original
science drivers lie in the astrophysical or heliophysical realms.

Some facilities that we hope will become active in the 2020s 
have the potential to bring us a significant jump in the number
of known, characterized comets.
The Rubin Observatory (c.f. \citealt{2009EM&P..105..101J}) and NEO Surveyor \citep{2021DPS....5330616M}, which will both be scanning
the skies for Solar System objects in the near future, will
provide us with the number statistics that would be incredibly helpful.
Estimates of the surveys' efficiencies 
suggest that we could be finding thousands
of new comets
\rev{\citep{2010Icar..205..605S,2020arXiv200907653V,2021plde.confE..56S}}. In particular, we can \rev{increase} the number of
LPCs that are discovered \rev{per year}, and the number of such comets that
are discovered \rev{beyond 5 and 10 au}, expanding baselines of behavior before activity has ramped up. This will also 
make it easier for follow-up observations to assess the 
properties of their nuclei. Another important consideration
that number statistics will help with is 
\rev{our} understanding \rev{of} the evolutionary paths
of the JFCs. 
For example, both surveys are supposed to be sensitive 
enough to sample the sub-kilometer JFC population
\rev{and we might expect such a population to exist as a result of mass loss and fragmentation over the comets' lifetimes. However it is also possible such small comets quickly disintegrate all the way to dust.}
\rev{Rubin and NEO Surveyor} will extend the number statistics into \rev{this } size regime. Hence the surveys will be better able to determine how well small comets survive their active lifetimes.

Finally, the two surveys will be very complementary to each other
since the combination of visible (reflected) and IR
(thermal) wavelength observations,
and observations spanning several years at 
multiple epochs, will be tremendously helpful for gauging 
gas, dust, and nucleus properties. In a similar vein as WISE/NEOWISE, NEO Surveyor will provide measurements of nucleus sizes, dust characteristics, and CO and CO$_2$ production of manifold larger statistical samples. 

The SPHEREx\footnote{Spectro-Photometer for the History of the Universe, Epoch of Reionization, and Ices Explorer.} mission  
will provide us with near-IR spectroscopic investigations
of dozens of comets. Importantly, such data will extend to
wavelengths where CO and CO$_2$ rovibrational bands emit, which
means SPHEREx may build upon results of
Akari with additional insight into
these important species \citep{2016arXiv160607039D}. 

Furthermore low-cost mission concepts could, if brought
to fruition, also 
address highly specific comet-related questions through a survey.
In particular cubesats can provide such survey work
for relatively low cost. For example
a small but fast, wide-field 
UV telescope in Earth-orbit could let us make measurements
of the OH electronic band near 309 nm in hundreds of cometary comae. 
Such a database, with observations covering all dynamical types,
and covering a range of heliocentric distances, and used in concert
with dust production measurements, 
could give a simple test of just how and when the water production is
tied to dust.
Cubesats could also be employed for an {\sl in situ} survey of
multiple comets. For example, equipping a fleet of
cubesats with replicas of the MIRO instrument
on Rosetta, and sending them out to a few dozen comets in
the inner Solar System, would
give us unprecedented views of cometary \rev{near-surface}
interiors and let us
assess how void space and consolidation evolve as a comet ages
through its active lifetime.

A survey with the James Webb Space Telescope (JWST) has the potential to let us take the next step in
our understanding of the comet population's nuclei, gas production,
and dust production. It will provide
more detailed gas production measurements, at a larger range of 
heliocentric distances, and for a broader range of species
(especially parent species), than ever
before. We will be able to watch the changing
release of various volatiles
over time as a comet approaches and recedes from the Sun, and do it
not just for exceptional comets such as Hale-Bopp, but for more typical
comets and comets from all dynamical classes and ages. Similar
synoptic coverage of the thermal and scattering properties of the
dust as the comet moves in its orbit will likewise let us investigate
how the grain properties change in response to the activity driver,
giving clues about the nature of the ice-rock mixture in cometary
subsurface layers.
This will come from not only spectroscopic assessment of the dust 
spectral energy distribution (and its resulting decomposition
into mineralogy) but 
also from resolved imaging of the dust coma. JWST's stable point-spread-function and
high spatial resolution will give us the best chance of overcoming
the coma confusion problem, and let us do so at a range of wavelengths,
thereby letting us have a better handle on overall thermal emission 
from the nucleus. Being able to do all this for 20 to 30 comets
would be spectacular.

There are of course many additional, large-scale space telescopes in
various levels of planning/concreteness that
would theoretically arrive in the decades of the 2030s and 2040s. The 
Roman Space Telescope, for example, will provide 
\rev{multiband red and near-IR imaging of cometary dust, as well as 
grism spectroscopy that covers the 1.5 micron 
water ice absorption band, thus potentially providing a survey of
icy grains} \citep{2018JATIS...4c4003H}. 
In the more distant future, observatories like the Large Ultraviolet Optical Infrared Surveyor (LUVOIR) or the Origins \rev{Space} Telescope (c.f. \citealt{NAP26141}) would further expand the samples. These observatories, combining significant improvements in sensitivity with high-resolution imaging and spectroscopy, may be used to get a statistical sense of the nature of low-level activity in comets at large heliocentric distances, as well as obtain large samples of surface constituents via spectroscopic studies, and explore the presence of other possible drivers of activity, like methane, at larger distances (cf. \citealt{2004come.book..317M} and \citealt{2000AJ....119..977B}). Such facilities, if capable of non-sidereal tracking,  would certainly provide a new jump in 
our understanding of comets by taking us to the next level
of detail on dozens to hundreds of these bodies. 

Radio-wavelength surveys with existing high-spatial
resolution facilities like ALMA
and with future facilities like \rev {the next generation Vary Large Array, ngVLA, and the Next Generation Arecibo Telescope (NGAT) }
\citep{2019BAAS...51g.244R} will provide great insight into
energetics of the gas coma. The NGAT would also
have a phased radar capability and thus provide a significantly
higher power output that has been previously possible. This would 
allow us to obtain more detailed pictures of the
large-grain (cm-scale) dust coma as well as the nucleus structure
on cm-scales. The number of radar-detected comets has
slowly increased since the predecessor volume, and is still
fairly small, so a boost
to the emitted power could drastically increase the number
of available comets that could be sampled in this way. 
Radio continuum measurements (in passive mode) at a variety
of sub-mm, mm, and cm wavelengths would let us sample
different depths in a nucleus, down to approximately a meter. This
has already been demonstrated with ALMA with observations of
Ganymede \citep{2021PSJ.....2....5D}. Again, an assessment of the
surface and sub-surface properties of a range of nuclei of varying
ages and activity levels could be insightful.

There are many excellent surveys of cometary properties
that are discussed throughout the current volume and the predecessor
volume. But often the statistics could be improved to sharpen
the conclusions by observing
more comets or by observing the same comets in more detail
(e.g. better temporal coverage, wavelength coverage, spatial resolution,
or spectral resolution). With growing data sets, selecting the necessary qualities of the data to undertake analyses of physical properties, and tailor selections to particular sub-populations, will require improved meta-data. Complex information models as the basis of meta-data, like PDS4 \citep{2021PSJ.....2..204R} \rev{and the Minor Planet Center's new Astrometry Data Exchange Standard (ADES; \citealt{2017ADES})}, which are associated with archived data, will facilitate these applications and analyses. The meta-data labels and automated tools will allow users to identify and extract the desired data for analysis. The question of when one has `enough' samples to draw a conclusion at a sufficient confidence level is not easy to answer ahead of time unless one has a good sense of the inherent diversity in the population.


\section{SUMMARY} 
\label{sec:summ}


Surveys have had a large impact on our knowledge base of comets. They provide a systematical approach towards collecting large samples of data reflective of cometary physical characteristics and behavior and more revealing of the comprehensive cometary object populations. 
It is important to also acknowledge that comet science can advance
as well with surveys done by telescopic assets that may have 
primarily astrophysical or heliophysical science drivers, and we
advocate for the continued use of such facilities. We also would
like to see that those facilities that are open to general observers
for targeted observations accommodate at least some of
the non-sidereal tracking
capability that is so important for Solar System science. 

Some key features in the sky survey analysis approaches include:
\begin{enumerate}
    \item most comets are now discovered by all-sky surveys, SLSs and TPSs. 
    \item the majority of yearly detections are also made by such surveys.
    \item critical margins remain outside survey coverage which create opportunities for targeted single-object observations and non-survey discovery of unique cometary objects. 
    \item the data from sky surveys are well-explored for discovery and astrometric measurements, while only cursorily exploited as a resource for physical and behavior characterization.
\end{enumerate}

For targeted larger-sample surveys:
\begin{enumerate}
    \item larger-sample sizes, the expanse of sample, and broadening of wavelength regimes will lead to more comprehensive understanding of compositional variations across orbit classes and sub-classes, and separate original composition from evolutionary effects.
    \item characterizing the physical states of comet components (e.g. gas phase, tail dust, nucleus spin, etc.) remains in the early stages of large-sample collection. 
\end{enumerate}

Larger-scale surveys in general will lead to improved earth-space situational awareness, understanding of the evolutionary processing of comets and solar system volatile transport, as well as solar system formation. Finally, future larger-scale surveys will require the use of advanced astro-informatic techniques and incorporation of AI routines to realistically process the increased data volume.

\vskip .5in
\noindent \textbf{Acknowledgments.} 

JMB and YRF acknowledge support from the NEO Surveyor mission, funded by
NASA Contract 80MSFC20C0045. YRF also acknowledges support from the Center for
Lunar and Asteroid Surface Science, funded by NASA Contract 80NSSC19M0214.
We acknowledge helpful discussions with A. Lovell that improved
this chapter.

\bibliographystyle{sss-full.bst}
\bibliography{refs.bib}

\end{document}